\documentclass[%
 reprint,
 amsmath,amssymb,
 aps,
]{revtex4-2}
\usepackage{mhchem}
\usepackage{gensymb}
\usepackage{graphicx}
\usepackage{dcolumn}
\usepackage{bm}

\begin{document}

\preprint{APS/123-QED}

\title{Superconductivity in the $\mathbb{Z}_2$ kagome metal KV$_3$Sb$_5$}%

\author{Brenden R. Ortiz}
 \email{ortiz.brendenr@gmail.com}
 \affiliation{Materials Department and California Nanosystems Institute, University of California Santa Barbara, Santa Barbara, CA, 93106, United States}%
 
\author{Eric M. Kenney}
 \affiliation{Physics Department, Boston College, Chestnut Hill, MA, 02467, United States}

\author{Paul M. Sarte}
 \affiliation{Materials Department and California Nanosystems Institute, University of California Santa Barbara, Santa Barbara, CA, 93106, United States}%
 
\author{Samuel M.L. Teicher}
 \affiliation{Materials Department, University of California Santa Barbara, Santa Barbara, CA, 93106, United States}%
 
\author{Ram Seshadri}
 \affiliation{Materials Department, University of California Santa Barbara, Santa Barbara, CA, 93106, United States}%

\author{Michael J. Graf}
 \affiliation{Physics Department, Boston College, Chestnut Hill, MA, 02467, United States}

\author{Stephen D. Wilson}
 \email{stephendwilson@ucsb.edu}
 \affiliation{Materials Department, University of California Santa Barbara, Santa Barbara, CA, 93106, United States}%

\date{\today}

\begin{abstract}
Here we report the observation of bulk superconductivity in single crystals of the two-dimensional kagome metal KV$_3$Sb$_5$. Magnetic susceptibility, resistivity, and heat capacity measurements reveal superconductivity below $T_c = 0.93$\,K, and density functional theory (DFT) calculations further characterize the normal state as a $\mathbb{Z}_2$ topological metal.  Our results demonstrate that the recent observation of superconductivity within the related kagome metal CsV$_3$Sb$_5$ is likely a common feature across the \textit{A}V$_3$Sb$_5$ (\textit{A}: K, Rb, Cs) family of compounds and establish them as a rich arena for studying the interplay between bulk superconductivity, topological surface states, and likely electronic density wave order in an exfoliable kagome lattice.
\end{abstract}

\maketitle

\section{Introduction}

The kagome lattice broadly supports an array of electronic phenomena and ground states of interest that span multiple fields within condensed matter physics. At the localized limit, the geometric frustration inherent to magnetic kagome lattices is the impetus for much of the historical work on these materials. Insulating kagome lattices, for example, are of interest as potential hosts of a quantum spin liquid state (e.g. Herbertsmithite, ZnCu$_3$(OH)Cl$_2$) \cite{freedman2010site,wulferding2010interplay,han2012refining,fu2015evidence,han2012fractionalized}. At the itinerant limit, kagome metals have also recently come into focus due to their potential to host novel correlated and topological electronic states. Even absent magnetic ordering, the kagome motif naturally gives rise to electronic structures with Dirac cones, flat bands, and the potential for topologically nontrivial surface states. Between these two extremes, kagome metals are predicted to support a wide array of instabilities, ranging from bond density wave order \cite{PhysRevB.87.115135,PhysRevLett.97.147202}, charge fractionalization \cite{PhysRevB.81.235115, PhysRevB.83.165118}, spin liquid states \cite{yan2011spin}, charge density waves \cite{PhysRevB.80.113102} and superconductivity \cite{PhysRevB.87.115135,ko2009doped}.

The \textit{A}V$_3$Sb$_5$ (\textit{A}: K, Rb Cs) compounds are a new model family of quasi two-dimensional kagome metals that were recently discovered~\cite{ortiz2019new}. These materials crystallize as layered materials in the $P6/mmm$ space group, forming layers of ideal kagome nets of V ions coordinated by Sb. The layers are separated by alkali metal ions, forming a highly two-dimensional, exfoliatable structure. Figure \ref{fig:Crystal} shows the structure of KV$_3$Sb$_5$ in top-down and isometric views. 

Since their discovery, \textit{A}V$_3$Sb$_5$ compounds have shown an array of interesting phenomena. Single crystals of KV$_3$Sb$_5$ were reported to exhibit an unusually large, unconventional anomalous Hall effect \cite{YangKV3Sb5science}. Both KV$_3$Sb$_5$ and CsV$_3$Sb$_5$ have also been investigated by angle-resolved photoemission spectroscopy (ARPES) and density-functional theory (DFT), demonstrating multiple Dirac points near the Fermi level \cite{ortiz2019new,YangKV3Sb5science,ortizCsV3Sb5}. The normal state of CsV$_3$Sb$_5$ can be described as a nonmagnetic, $\mathbb{Z}_2$ topological metal \cite{ortizCsV3Sb5} and topologically-protected surface states have been identified close to the Fermi level.

One widely sought electronic instability on a two-dimensional kagome lattice is the formation of superconductivity.  Layered kagome metals that superconduct are rare, and the interplay between the nontrivial band topology and the formation of a superconducting ground state makes this a particularly appealing space for realizing unconventional quasiparticles. In the localized electron limit, superconductivity competes with a variety of other electronic instabilities at different fillings~\cite{PhysRevLett.110.126405,wen2010interaction}.  In the itinerant limit, unconventional superconductivity is also predicted to emerge via nesting-driven interactions~\cite{PhysRevB.85.144402}.  This mechanism, analogous to theories for doped graphene (which shares the hexagonal symmetry of the kagome lattice) \cite{nandkishore2012interplay,nandkishore2012chiral}, relies upon scattering between saddle points of a band at the M points of the two-dimensional Brillouin zone.    

CsV$_3$Sb$_5$ was recently proposed as the first material potentially hosting a nesting-driven superconducting state, following the observation of bulk superconductivity below $T_c=2.5$\,K in both single crystals and powders ~\cite{ortizCsV3Sb5}.  However, the presence of superconductivity in the remaining compounds in the \textit{A}V$_3$Sb$_5$ class of materials, which share nearly identical electronic structures, remains an open question. Although superconductivity was not initially observed in KV$_3$Sb$_5$ or RbV$_3$Sb$_5$ above 1.8\,K, the impact of sample quality (particularly with regards to A-site occupancy) and of insufficiently low temperature measurements remain unaddressed.

In this work, we report the discovery of bulk superconductivity in single crystals of KV$_3$Sb$_5$ below $T_c = 0.93$\,K. We characterize the superconducting state using a combination of low temperature susceptibility, heat capacity, and electrical resistivity measurements on stoichiometric samples where K vacancies have been minimized. Consistent with prior work on CsV$_3$Sb$_5$ \cite{ortizCsV3Sb5}, we also present density-functional theory (DFT) results showing that the normal state of KV$_3$Sb$_5$ is also characterized as a $\mathbb{Z}_2$ topological metal. Our data demonstrate that $\mathbb{Z}_2$ topology and superconducting ground states are common properties of the new class of \textit{A}V$_3$Sb$_5$ kagome metals. 

\section{Methods}
\subsection{Synthesis}
Single crystals of KV$_3$Sb$_5$ were synthesized from K (solid, Alfa 99.95\%), V (powder, Sigma 99.9\%) and Sb (shot, Alfa 99.999\%). As-received vanadium powder was purified using EtOH and concentrated HCl to remove residual oxides. Powder preparation was performed within an argon glove box with oxygen and moisture levels $<$0.5\,ppm. Single crystals of KV$_3$Sb$_5$ were then synthesized using a self-flux method using an eutectic mixture of KSb$_2$ and KSb \cite{sangster1993k}, mixed with VSb$_2$. Elemental reagents were initially milled in a sealed, pre-seasoned tungsten carbide vial to form the precursor composition, which is approximately 50\,at.\% K$_{x}$Sb$_{y}$ eutectic and approximately 50\,at.\% VSb$_2$. The precursor powder was subsequently loaded into alumina crucibles and sealed within stainless steel jackets. The samples were heated to 1000\degree C at 250\degree C/hr and soaked there for 24\,h. The mixture was subsequently cooled to 900\degree C at 100\degree C/hr and then further to 400\degree C at 2\degree C/hr. Once cooled, the flux boule is crushed, and crystals were extracted mechanically. Mechanical extraction and elimination of aqueous solvents during crystal extraction appear essential to recovering fully occupation of the potassium site. Crystals are hexagonal flakes with brilliant metallic luster. Samples are routinely 1-3\,mm in side length and 0.1-0.5\,mm thick. Crystals are naturally exfoliatable and readily cleave along the \textit{a-b} plane, consistent with the layered structure.  Elemental composition of the crystals was assessed using energy dispersive x-ray spectroscopy (EDX) using a APREO C scanning electron microscope.

\subsection{Susceptibility, electrical transport, and heat capacity measurements}

Magnetometry data between 300\,K and 2\,K were collected using a Quantum Design Squid Magnetometer (MPMS3) in vibrating-sample measurement mode (VSM). Samples were mounted on a quartz paddle using GE varnish with crystals aligned with the c-axis perpendicular to the field. Alternating current (AC) magnetic susceptibility measurements between 3.5\,K and 300\,mK were performed in a $^{3}$He refrigerator using a custom susceptometer consisting a pair of astatically wound pickup coils inside a drive coil. Samples were mounted on a sapphire rod with silver paint and thermally anchored to the refrigerator via a gold wire. For the AC measurements, the c-axis was also aligned parallel to the field, and data were collected at a frequency of 711.4\,Hz. The susceptometer response was calibrated against standard samples measured in a Quantum Design AC susceptometer.  Due to the orientation of the crystals, a correction for demagnetization was applied in the quantitative analysis of the superconducting volume fraction. The crystals (hexagonal platelets) were approximated as cylindrical volumes, though calibrations were also performed using similarly sized samples of Ta to accurately correct for any systematic offsets in the data.

Heat capacity and electrical resistivity measurements were performed using a Quantum Design 14~T Dynacool Physical Property Measurement System (PPMS) with a $^{3}$He/$^{4}$He dilution refrigerator insert option. Apezion N grease was used to ensure thermal and mechanical contact with heat capacity stage. Crystals were mounted such that the c-axis was parallel to the applied H-field.  Electrical contacts were made in a standard 4-point geometry using gold wire and silver paint. Thermal contact and electrical isolation was ensured using layers of GE varnish and cigarette paper. Samples were again mounted such that the c-axis was parallel to the field (flat plates mounted flush on resistivity stage). An alternating current of 1.25\,mA and 12.2\,Hz was driven in the $ab$-plane.

\subsection{Band structure calculations}
In order to address potential orbital localization (particularly V $d$) and the van der Waals forces that play a role in $c$-layer stacking in this compound, we performed simulations using the SCAN functional \cite{Sun2015} and D2/D3 \cite{Grimme2006,Grimme2011} dispersion corrections as well as the more standard PBE functional \cite{Perdew1996}. Calculations described in the text employed PBE with D3 corrections as this functional choice and gave relaxed lattice parameters with excellent matching to the room temperature experimental values for KV$_3$Sb$_5$ (see Table SFig 1). Irrep assignments (see Table SFig. 2) for the $\mathbb{Z}_2$ topological invariant calculation were performed using the \textsc{Irvsp} program in conjunction with VASP \cite{gao2020irvsp, ESI}.

\section{Experimental Results}

\begin{figure}
\includegraphics[width=\columnwidth]{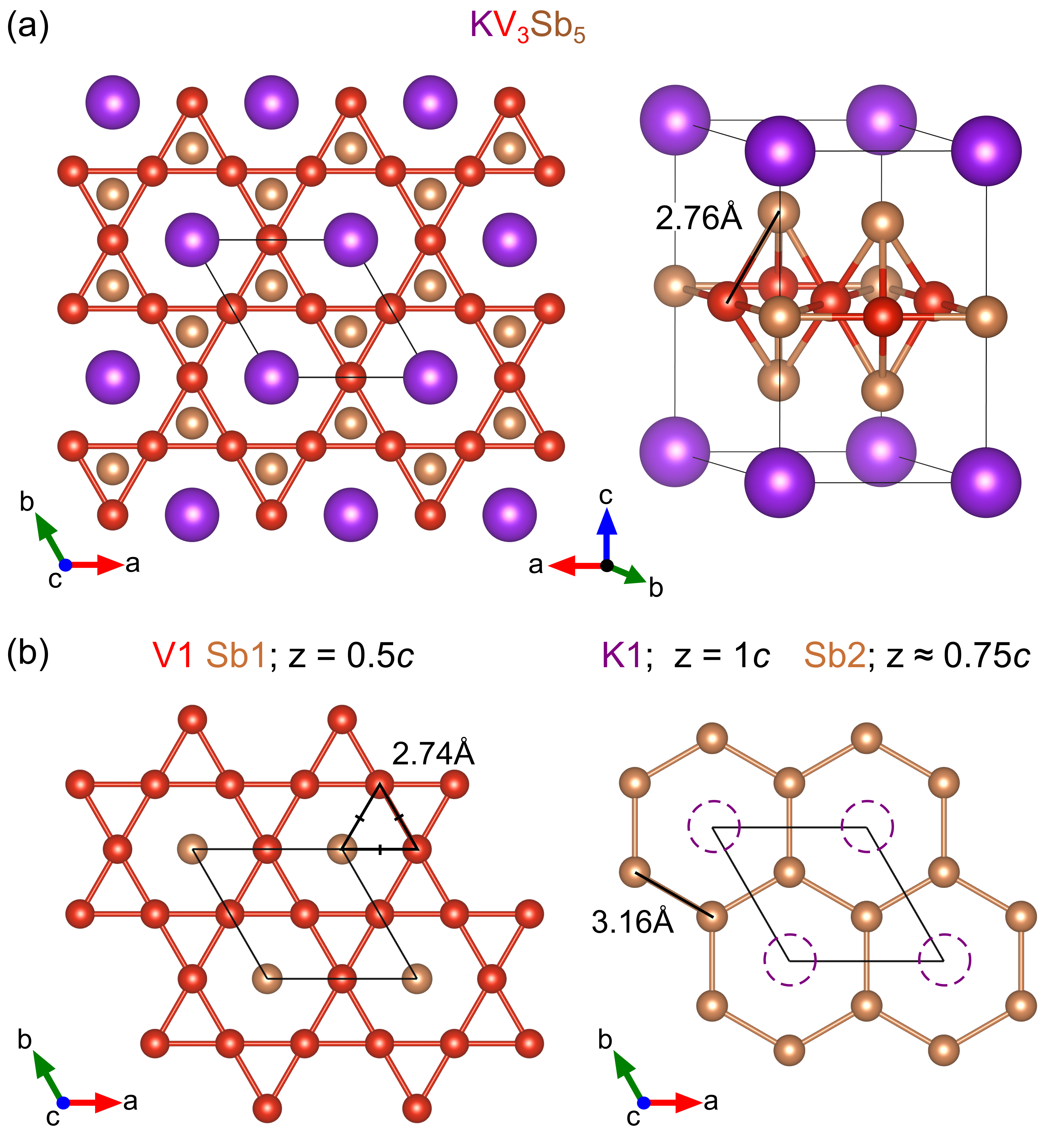}
\caption{Crystal structure of KV$_3$Sb$_5$, the prototype structure for the \textit{A}V$_3$Sb$_5$ family of kagome metals. (a) The kagome lattice of vanadium is immediately apparent from the top-down view of the structure. Bond lengths $<$3\AA\, are drawn in the isometric perspective to highlight the layered nature of KV$_3$Sb$_5$. (b) Intuitive decomposition of the structure highlights the kagome lattice and graphite-like (antimonene) layers.}
\label{fig:Crystal}
\end{figure}

\begin{figure*}[t]
\centering
\includegraphics[width=7.05in]{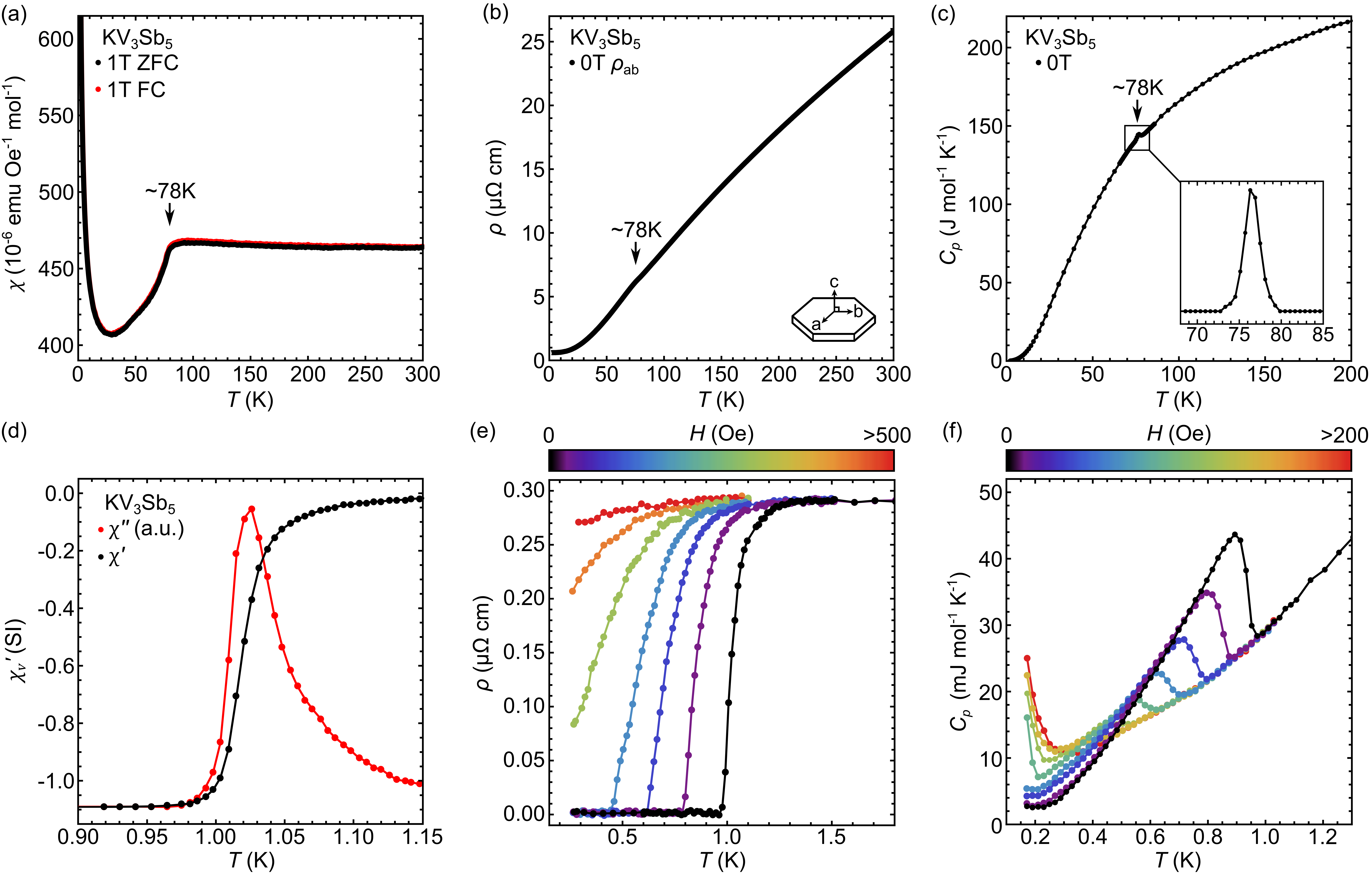}
\caption{(a,b,c) Susceptibility, electrical resistivity, and heat capacity showing behavior of stoichiometric single crystals above 2\,K. All measurements show an anomaly at 78\,K, coinciding with emergence of a charge density wave. Magnetization results indicate that KV$_3$Sb$_5$ is a Pauli paramagnet at high temperatures. As shown previously, the weak Curie tail at low temperature can be fit with a small concentration of impurity spins. Resistivity is low, indicating a high mobility metal. The lambda-like anomaly in the heat capacity is shown, magnified, with a spline interpolation used to isolate the transition (d,e,f) Susceptibility, electrical resistivity, and heat capacity measurements below 2\,K highlight the onset of bulk superconductivity in KV$_3$Sb$_5$. A well-defined Meissner state is observed in susceptibility, which coincides with the zero-resistivity state and a sharp heat-capacity anomaly.}
\label{fig:Transport}
\end{figure*}

\subsection{Structure and Composition}
The \textit{A}V$_3$Sb$_5$ family of kagome metals are layered, exfoliable materials consisting of V$_3$Sb$_5$ slabs intercalated by alkali metal cations. As shown in Figure \ref{fig:Crystal}(a), the kagome network of vanadium is immediately obvious, as is the layered nature of the crystal structure (all bonds $<$3\AA\, are shown in isometric perspective). For a conceptual interpretation of the V$_3$Sb$_5$ network, we can decompose the structure further into the individual sublattices. Figure \ref{fig:Crystal}(b) shows the V$_3$Sb$_5$ slab decomposed into the sublattices formed by the V1, Sb1, and Sb2 atomic sites. The V1 sublattice forms the kagome lattice, and is interpenetrated with the Sb1 sublattice. The Sb2 sublattice encapsulates the kagome layer with a graphite-like (antimonene) layer of antimony. The alkali metals fill the natural space left between the antimonene layers. We note that the V-V bond distances are short, 2.74\AA\, , in KV$_3$Sb$_5$.

While KV$_3$Sb$_5$ is nearly identical to CsV$_3$Sb$_5$ in structure, the alkali K layer is more easily deintercalated from the lattice and the material has a tendency to form a variety of sub-stoichiometric compositions. In our prior report, as-grown crystals were slightly deintercalated (approximately K$_{0.9}$V$_3$Sb$_5$) \cite{ortiz2019new}. We ascribe the potassium loss to the flux removal process in that report. To highlight the ease with which potassium is lost, we explored a variety of postgrowth treatments (e.g. hydrothermal, mineral acid etches, vapor transport). We estimate that the as-grown crystals can be processed to remove approximately 25-30\% of the potassium. For instance, crystals with compositions of K$_{0.7}$V$_3$Sb$_5$ have been formed by sealing as-grown crystals with small quantities of PtCl$_2$ and allowing the halogen to react with the potassium at elevated temperatures. A ``reverse hydrothermal'' method was also explored, by sealing crystals in glass tubes with purified water and heating to 150\degree C. This method also reduces the potassium content to K$_{0.7}$V$_3$Sb$_5$. Beyond this limit however, the material begins to decompose into VSb$_2$. Even within the allowable range of deintercalation, crystals of K$_{1-x}$V$_3$Sb$_5$ become progressively more brittle as potassium removal progresses.

While the tunability of the potassium site is an interesting degree of freedom when studying transport in KV$_3$Sb$_5$, the focus in this paper is on \textit{stoichiometric} crystals with fully occupied K1 sites. The discovery of superconductivity in stoichiometric crystals of CsV$_3$Sb$_5$ highlights the need to control the alkali-metal occupancy. Stoichiometric crystals ($\sim$11.1 at.\% K as measured by energy dispersive spectroscopy) were produced following the new methods outlined above with the main consideration being avoiding all contact with water. Mild solvents like isopropyl alcohol do not seem to affect the potassium content of the crystals. 

\subsection{Normal state transport, susceptibility, and heat capacity results}

Figure \ref{fig:Transport} presents a suite of magnetization, electrical resistivity, and heat capacity measurements of single crystals of KV$_3$Sb$_5$. Panels (a-c) present data characterizing the normal state of KV$_3$Sb$_5$ above 2\,K. In all three measurements, data reveal an anomaly at $T^*=78$\,K, consistent with prior work \cite{ortiz2019new}. This $T^*$ feature is shared across the \textit{A}V$_3$Sb$_5$ family, though the onset temperature varies: 78-80\,K in KV$_3$Sb$_5$, 93\,K in CsV$_3$Sb$_5$, 110\,K in RbV$_3$Sb$_5$.

Specifically, normal state magnetization data (Figure \ref{fig:Transport}(a)) indicate temperature-independent Pauli paramagnetism above $T^*$.  Below $T^*$, there is an abrupt step-like decrease in susceptibility, suggestive of a partial gapping of the Fermi surface.  The same $T^*$ transition also appears as a weak inflection in resistivity data presented in Figure \ref{fig:Transport}(b) and as a weak $\lambda$-like anomaly in heat capacity data in Figure \ref{fig:Transport}(c). The influence of the anomaly in both heat capacity and resistivity is weaker than in previous single crystal studies of CsV$_3$Sb$_5$ \cite{ortizCsV3Sb5}, and this distinction is also present in prior measurements on powders  \cite{ortiz2019new, ortizCsV3Sb5}. 

Recent studies of CsV$_3$Sb$_5$ further observed the onset of weak structural superlattice peaks below $T^*$, suggesting that a structural transition onsets as a secondary response to a primary electronic order parameter \cite{ortizCsV3Sb5}. This mechanism and the presence of a charge density wavelike instability is the likely source of the $T^*$ anomaly throughout the AV$_3$Sb$_5$ family of materials. Recently, scanning tunneling microscopy (STM) has confirmed the presence of the charge density wave in KV$_3$Sb$_5$.\cite{yuxiaoKVS}

\subsection{Low-temperature transport, susceptibility, and heat capacity measurements}

Panels (d-f) in the bottom row of Figure \ref{fig:Transport} present data characterizing KV$_3$Sb$_5$ single crystals below 2\,K. All probes reveal that KV$_3$Sb$_5$ crystals are bulk superconductors with $T_c \approx 0.93$\,K. Figure \ref{fig:Transport} (d) presents the real and imaginary components of the AC magnetic susceptibility showing bulk superconductivity with a well-defined Meissner state and a superconducting volume fraction of $4\pi\chi_v \approx -1.1$. We estimate the error in $\chi_v$ is on the order of 10-15\% due to the approximations used when correcting for the sample's demagnetization. The imaginary component $\chi^{\prime \prime}$ is shown in arbitrary units to highlight the onset of the transition.  

Figure \ref{fig:Transport} (e) presents low-temperature resistivity measurements similarly exhibiting a sharp superconducting phase transition into a zero-resistivity state. The superconducting state is quenched quickly by the application of a c-axis aligned magnetic field, and the zero-resistivity condition is suppressed by $\mu_0 H=300$\,Oe. Crystals of KV$_3$Sb$_5$ have inherently low resistivity in the normal state ($\approx0.3$ $\mu\Omega$-cm at 2\,K), comparable to measurements of stoichiometric CsV$_3$Sb$_5$.\cite{ortizCsV3Sb5} To ensure good noise-to-signal ratio, a relatively large ($\sim$1\,mA) AC current was used. We don't see any deleterious effect of the large current on the superconducting temperature at this time.

Low-temperature heat capacity measurements shown in \ref{fig:Transport} (f) also indicate a well-defined entropy anomaly at $T_c$. Combined with the magnetization and resistivity measurements, stoichiometric KV$_3$Sb$_5$ can be unambiguously classified as a bulk superconductor. We note that heat capacity measurements also indicate the presence of a nuclear Schottky anomaly at the lowest temperatures measured. The anomaly matches known contributions from \ce{^{51}_{}V} nuclei observed in other vanadium-containing compounds \cite{andresV2O3}.

\begin{figure}
\includegraphics[width=\columnwidth]{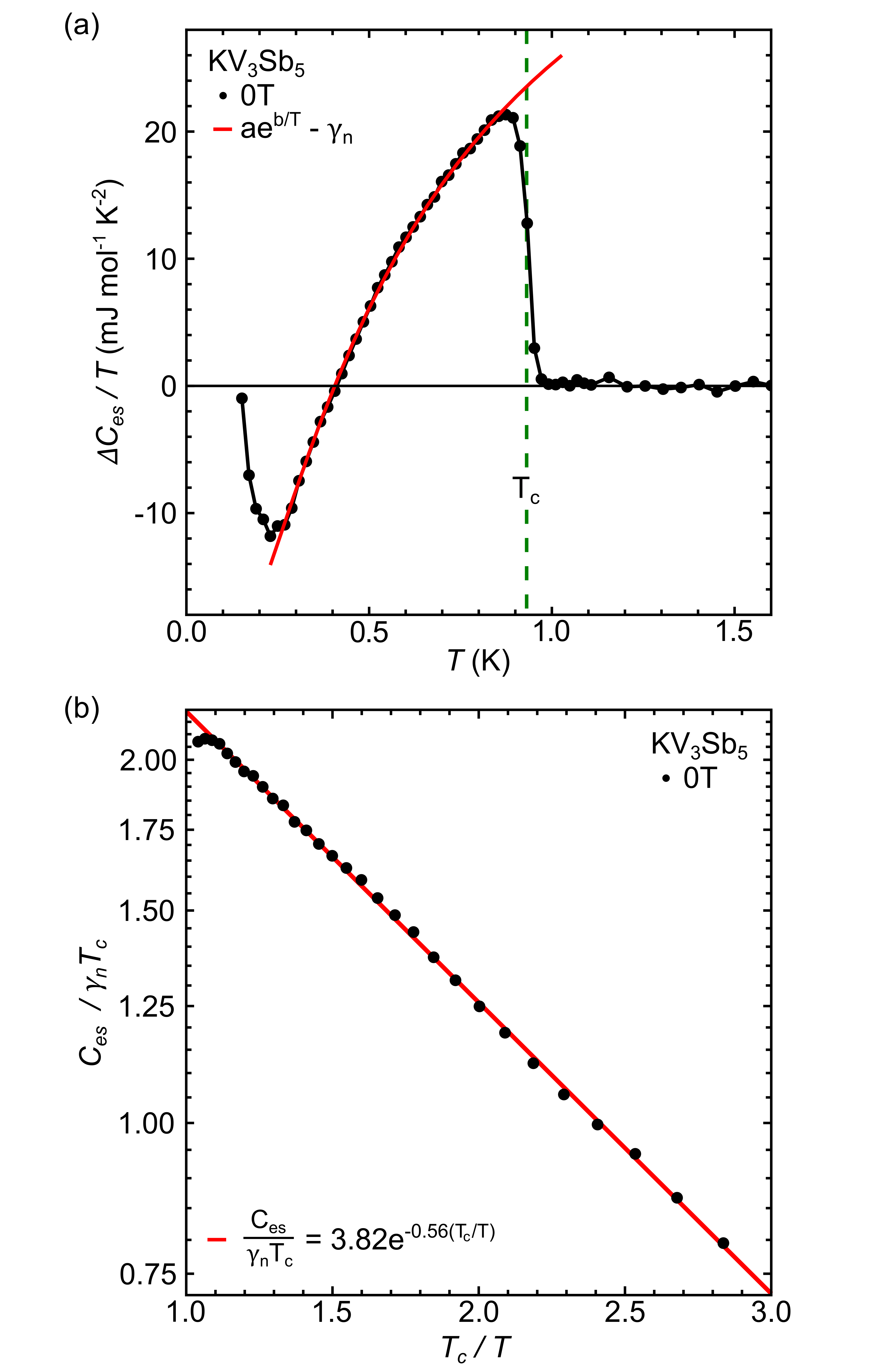}
\caption{(a) The specific heat capacity difference $\Delta C_{es}$ between the superconducting and normal states ($\Delta C_{es} =  C_p-C_{p,3T}$) for KV$_3$Sb$_5$.  Here, $\gamma_n = 22.8$\,mJ\,mol$^{-1}$\,K$^{-1}$ and $\theta_D = 141$\,K have been modeled from the 3\,T data set. (b) The electronic heat capacity $C_{es}=C_{p,0T}-AT^3$ below the transition (and above the Schottky anomaly) vanishes exponentially with temperature.  Red lines show the exponential fits to the data as described in the text.}
\label{fig:Entropy}
\end{figure}

The normal state heat capacity data ($T > T_c$) fits to the form $C_{p} = \gamma_n T + AT^3$  \cite{ESI}, allowing $\gamma_n = 22.8$\,mJ\,mol$^{-1}$\,K$^{-2}$ to be determined. From $A$, we estimate that the Debye temperature for KV$_3$Sb$_5$ is approximately $\theta_D = 141$\,K. The overall trends are consistent with prior high temperature $C_p$ data published on powders \cite{ortiz2019new}. Removing the lattice contribution to the heat capacity allows us to examine the electronic specific heat ($C_{es} = C_{p,0T} - AT^3$) in detail. Further, we can isolate the superconducting transition ($\Delta C_{es} = C_{p,0T} - C_{p,3T}$) by removing the normal state specific heat collected under a 3\,T field.

Figure \ref{fig:Entropy} (a) shows the heat capacity associated with the onset of superconductivity as $\Delta C_{es}/T$. The lattice component was verified from the normal state data collected under a 3\,T field. The red curve is a fit to the empirical, exponential approximation of the BCS formalism, using the form $C_{es}/T \approx ae^{b/T}$ ($\Delta C_{es}/T \approx ae^{b/T} - \gamma_n$). This curve is used to help determine $\Delta C_{es}/\gamma_n T_c$ and $T_c = 0.93$\,K, where the dimensionless specific jump at $T_c$ is estimated to be $\Delta C_{es}/\gamma_n T_c = 1.03$ and the crossover point in $\Delta C_{es}$ occurs at 0.44$T_c$. Both values are quite small relative to the BCS weak coupling limit $\Delta C_{es}/\gamma_n T_c \approx 1.43$ and 0.6$T_c$, respectively.

From Figure \ref{fig:Entropy}, the data below $T_c$ vanishes exponentially with temperature.  Figure \ref{fig:Entropy} (b) shows the experimental data below $T_c$, modeled with $C_{es} = ae^{-b(T_c/T)}$.  Again, the fit value of $b=0.56$ is substantially smaller than the expectations of BCS theory ($b_{BCS}=1.44$) suggesting a deviation from the traditional weak-coupling theory with an isotropic gap. Unfortunately, interference from the Schottky anomaly at low temperature precludes a more exhaustive examination of the superconducting gap and the field dependence of $\gamma_n$ from the present $C_p$ data.  

\subsection{$\mathbb{Z}_2$ Evaluation}

In this section, we determine whether KV$_3$Sb$_5$ shares the same topological classification as its isostructural variant CsV$_3$Sb$_5$ . Unlike many of the more heavily studied kagome lattices (e.g. Fe$_3$Sn$_2$,\cite{ye2018massive,wang2020giant} Mn$_3$Ge,\cite{nayak2016large,kiyohara2016giant} Co$_3$Sn$_2$S$_2$\cite{vaqueiro2009powder,xu2016topological,morali2019fermi}, etc.), \textit{A}V$_3$Sb$_5$ compounds possesses both time-reversal and inversion symmetry. The $\mathbb{Z}_2$ topological invariant between each pair of bands near the Fermi level can be calculated by simply analyzing the parity of the wave function at the TRIM (time-reversal invariant momentum) points \cite{fu2007topological}. 

\begin{figure}
\includegraphics[width=\columnwidth]{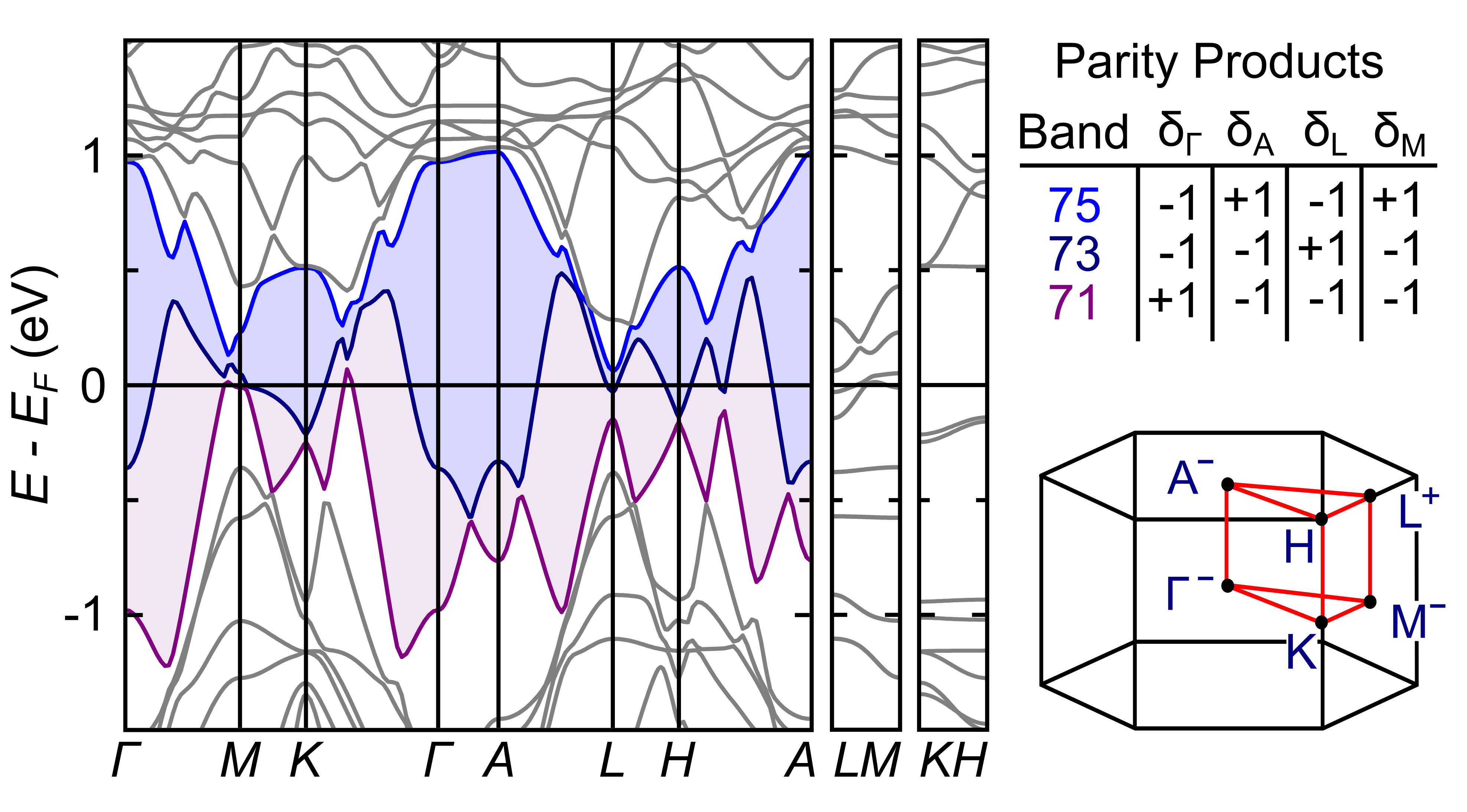}
\caption{DFT electronic band structure for KV$_3$Sb$_5$ near the Fermi level. The continuous gap (shaded) is noted, with band indices and parity products given. Note that surface states at the $\overline{M}$ are expected to be topologically protected, in analogy to CsV$_3$Sb$_5$.}
\label{fig:Band}
\end{figure}

Figure \ref{fig:Band} shows the calculated electronic structure of KV$_3$Sb$_5$. The symmetry-enforced direct gap between the bands near $E_F$ has been highlighted. As with CsV$_3$Sb$_5$, topologically nontrivial surface state crossings at the $\overline{M}$ TRIM point (between bands 71 and 73) are expected. The combination of topologically nontrivial surface states near $E_F$, combined with the continuous direct gap allow us to classify KV$_3$Sb$_5$ as a $\mathbb{Z}_2$ topological metal. 

\section{Discussion}
The observation of superconductivity in both KV$_3$Sb$_5$ and CsV$_3$Sb$_5$ suggests that the known members of the \textit{A}V$_3$Sb$_5$ (A: K, Rb, Cs) kagome metals may all be $\mathbb{Z}_2$ topological metals with superconducting ground states. All compounds also exhibit very similar $T^*$ anomalies, which are suggestive of appreciable electronic interactions driving an accompanying charge or bond density wave order. Based on our observations in KV$_3$Sb$_5$, the emergence of superconductivity may depend strongly on the alkali-metal site occupancy.

Long-range magnetic ordering is not observed in KV$_3$Sb$_5$ from previous neutron diffraction measurements \cite{ortiz2019new}, and more recent muon spin resonance ($\mu$SR) measurements performed on near-stoichiometric polycrystalline powders of KV$_3$Sb$_5$ observe no signature of localized, electronic magnetic moments \cite{2020grafKV3Sb5}. Nevertheless, recent reports have shown unconventional transport phenomena in substoichiometric K$_{1-x}$V$_3$Sb$_5$ crystals attributed to the presence of local, disordered magnetic moments \cite{2020MazProximitized, YangKV3Sb5science}. Direct experimental detection of the proposed magnetic defects in K$_{1-x}$V$_3$Sb$_5$, however, remains an open challenge. 

Heat capacity measurements suggest a superconducting state that deviates from the BCS expectations of single-band, weak coupling s-wave superconductivity.  The small $\Delta C_{es}/\gamma_n T_c=1.03$ value potentially indicates a nonuniform gap structure in momentum space, or alternatively the possibility of a multi-band gap structure \cite{johnston2013elaboration}.  Single band d-wave superconductivity nominally has a value of $\Delta C_{es}/\gamma_n T_c=0.95$, which is not far from the observed value. The low temperature Schottcky anomaly however currently precludes canonical measurements of the field dependence of $\gamma_n$ to assess the presence of nodes within the gap structure.  Future measurements via low-temperature scanning tunneling microscopy or angle-resolved photoemission measurements exploring this possibility are highly desired.

In comparison to prior results on K$_{0.92}$V$_3$Sb$_5$ and K$_{0.85}$V$_3$Sb$_5$ crystals \cite{ortiz2019new}, the electrical resistivity collected from stoichiometric KV$_3$Sb$_5$ (Figure \ref{fig:Transport}(a)) shows marked changes. Stoichiometric samples exhibit significantly reduced resistivity values, with the residual resistivity at 2\,K ($\rho_0$) dropping nearly two orders of magnitude from K$_{0.85}$ ($\approx$300$\mu\Omega$-cm), K$_{0.92}$ ($\approx$10$\mu\Omega$-cm), and K$_{1.0}$ ($\approx$1$\mu\Omega$-cm). At the stoichiometric limit, KV$_3$Sb$_5$ is an astonishingly good metal, considering the relatively complex crystal structure.

The strong influence of stoichiometry and disorder on the normal state electrical transport of KV$_3$Sb$_5$ is also reflected in the stability of its superconducting state. The lack of superconductivity in powders of KV$_3$Sb$_5$ reported previously \cite{ortiz2019new} is most reasonably attributed to the off-stoichiometry of potassium. Concurrent studies on potassium deficient K$_{0.7}$V$_3$Sb$_5$ crystals have shown they are not natively superconducting, although the possibility for proximitized superconductivity remains \cite{2020MazProximitized}.  Investigating the underlying connection between potassium occupancy and superconductivity in KV$_3$Sb$_5$ as well as the connection between $T^*$ and $T_c$ across the \textit{A}V$_3$Sb$_5$ family are interesting avenues for future study.

\section{Conclusion}

Our results have demonstrated the synthesis of high-purity, stoichiometric single crystals of KV$_3$Sb$_5$. Stoichiometric crystals exhibit bulk superconductivity with $T_c = 0.93$\,K and have a substantially reduced normal state residual resistivity. Similar to CsV$_3$Sb$_5$, DFT results demonstrate that KV$_3$Sb$_5$ can be classified as a $\mathbb{Z}_2$ topological metal with a number of non-trivial surface crossings near the Fermi level. Our results suggest that superconductivity is common across the \textit{A}V$_3$Sb$_5$ family of quasi two-dimensional kagome metals. They further motivate the continued exploration of \textit{A}V$_3$Sb$_5$ as materials supporting a rich interplay between superconductivity, non-trivial band topology, and potential correlation physics.

\section{Acknowledgments}
This work was supported by the National Science Foundation (NSF) through Enabling Quantum Leap: Convergent Accelerated Discovery Foundries for Quantum Materials Science, Engineering and Information (Q-AMASE-i): Quantum Foundry at UC Santa Barbara (DMR-1906325). The research made use of the shared facilities of the NSF Materials Research Science and Engineering Center at UC Santa Barbara (DMR- 1720256). The UC Santa Barbara MRSEC is a member of the Materials Research Facilities Network. (www.mrfn.org). B.R.O. and P.M.S. also acknowledge support from the California NanoSystems Institute through the Elings Fellowship program. S.M.L.T has been supported by the National Science Foundation Graduate Research Fellowship Program under Grant No. DGE-1650114

\bibliography{KV3Sb5_SC}

\end{document}